\documentstyle[11pt]{article}

\begin{document}

\input{epsf}

\def\beq{\begin{equation}}
\def\eeq{\end{equation}}
\def\bea{\begin{eqnarray}}
\def\eea{\end{eqnarray}}
\def\beas{\begin{eqnarray*}}
\def\eeas{\end{eqnarray*}}
\def\ov{\overline}
\def\ot{\otimes}

\newcommand{\hf}{\mbox{$\frac{1}{2}$}}
\def\sig{\sigma}
\def\De{\Delta}
\def\af{\alpha}
\def\be{\beta}
\def\la{\lambda}
\def\ga{\gamma}
\def\ep{\epsilon}
\def\vep{\varepsilon}
\def\half{\frac{1}{2}}
\def\third{\frac{1}{3}}
\def\fth{\frac{1}{4}}
\def\sth{\frac{1}{6}}
\def\tth{\frac{1}{24}}
\def\tde{\frac{3}{2}}

\def\zb{{\bar z}} 
\def\psib{{\bar \psi}} 
\def\etab{{\bar \eta }}
\def\gab{{\bar \ga}}
\def\vev#1{\langle #1 \rangle}
\def\inv#1{{1 \over #1}}

\def\CA{{\cal A}}       \def\CB{{\cal B}}       \def\CC{{\cal C}}
\def\CD{{\cal D}}       \def\CE{{\cal E}}       \def\CF{{\cal F}}
\def\CG{{\cal G}}       \def\CH{{\cal H}}       \def\CI{{\cal J}}
\def\CJ{{\cal J}}       \def\CK{{\cal K}}       \def\CL{{\cal L}}
\def\CM{{\cal M}}       \def\CN{{\cal N}}       \def\CO{{\cal O}}
\def\CP{{\cal P}}       \def\CQ{{\cal Q}}       \def\CR{{\cal R}}
\def\CS{{\cal S}}       \def\CT{{\cal T}}       \def\CU{{\cal U}}
\def\CV{{\cal V}}       \def\CW{{\cal W}}       \def\CX{{\cal X}}
\def\CY{{\cal Y}}       \def\CZ{{\cal Z}}

\newcommand{\np}{Nucl. Phys.}
\newcommand{\pl}{Phys. Lett.}
\newcommand{\prl}{Phys. Rev. Lett.}
\newcommand{\cmp}{Commun. Math. Phys.}
\newcommand{\jmp}{J. Math. Phys.}
\newcommand{\jpamg}{J. Phys. {\bf A}: Math. Gen.}
\newcommand{\lmp}{Lett. Math. Phys.}
\newcommand{\ptp}{Prog. Theor. Phys.}

\newif\ifbbB\bbBfalse                
\bbBtrue                             

\ifbbB   
 \message{If you do not have msbm (blackboard bold) fonts,}
 \message{change the option at the top of the text file.}
 \font\blackboard=msbm10 
 \font\blackboards=msbm7 \font\blackboardss=msbm5
 \newfam\black \textfont\black=\blackboard
 \scriptfont\black=\blackboards \scriptscriptfont\black=\blackboardss
 \def\Bbb#1{{\fam\black\relax#1}}
\else
 \def\Bbb{\bf}
\fi

\def\id{{1\! \! 1 }}
\def\bo{{\Bbb 1}}
\def\bI{{\Bbb I}}
\def\bC{{\Bbb C}} 
\def\bZ{{\Bbb Z}}
\def\CN{{\cal N}}

\title{The XXC Models}
\author{{\bf Z. Maassarani}\thanks{Work supported by NSERC 
(Canada) and FCAR (Qu\'ebec).} \\
\\
{\small D\'epartement de Physique, Pav. A-Vachon}\\
{\small Universit\'e Laval,  Ste Foy, Qc,  
G1K 7P4 Canada}\thanks{email address: zmaassar@phy.ulaval.ca} \\}
\date{}
\maketitle

\begin{abstract}
A class of recently introduced multi-states XX models  
is generalized to include a deformation parameter. This 
corresponds to an additional nearest-neighbor $CC$ interaction in 
the  defining quadratic hamiltonian.
Complete integrability of the one-dimensional models is 
shown in the context of the quantum inverse scattering method.
The new $R$-matrix is derived. 
The diagonalization of the XXC models is carried 
out using the algebraic Bethe Ansatz. 
 
\end{abstract}
\vspace{5cm}
\noindent
\hspace{1cm} December  $8^{\rm th}$ 1997\hfill\\
\hspace*{1cm} LAVAL-PHY-27/97\hfill\\
\hspace*{1cm} solv-int/9712008

\thispagestyle{empty}

\newpage

\setcounter{page}{1}

\section{Introduction}

Multi-states bosonic  Hubbard 
models which generalize  the usual $su(2)$ model were recently introduced. 
These models were shown to be integrable 
and to have an extended symmetry \cite{zm1,zm2,ff}.  
They are built by coupling two copies of
XX `free-fermions' models \cite{sh123,mm,ff}. These higher dimensional 
`fermion' models were not obtained as part of a quantum group structure.
The quantum group approach to the construction of solutions to the
Yang-Baxter equation results in higher dimensional models with symmetries
reflecting the root structure of the underlying algebra \cite{jiba}. 
The combination lowering-raising operators 
in the $R$-matrix generically involves all the roots of the algebra. 
The `free-fermions' models are also higher dimensional and they  have
simple expressions in terms of the generators of the 
fundamental representations of $su(n)$. However they involve
only a subset of the set of roots of $su(n)$. The diagonalization of the models 
reveals an $su(2)$ structure which does not correspond to the known
integrable higher-spin representations.
 
An integrable deformation of the XX models is found and studied in this
paper. For the lowest dimensional
model the XXC model is nothing but the usual $su(2)$ XXZ model.
The $R$-matrix for the higher dimensional models retains the form of 
the XXZ model and can be seen as a multistate generalization
of the latter model. The matrix is given and shown to satisfy 
the Yang-Baxter equation. Specific representations are given.
Integrability of the models is a simple consequence of the 
quantum inverse scattering framework. 
The symmetries of the new models are obtained. 
Diagonalization of  the XXC  models
using the algebraic Bethe Ansatz method shows that their spectrum 
is highly degenerate, and is a generalization of the $su(2)$ XXZ spectrum.
I conclude with some remarks. 

\section{The XXC models}\label{xxcmod}

Consider a  matrix $\check{R}$ of  the form
\beq
\check{R}(\la;\ga)=  P^{(1)}\sin\ga +P^{(2)} \sin(\ga-\la) + P^{(3)} 
\sin \la\label{rcm}
\eeq
where $\ga$ is an arbitrary complex parameter and $\la$ is 
the spectral parameter.
The operators $P^{(i)}$ act on the tensor product of two copies 
of the same space $\bC^n$. 
They are taken to satisfy the algebra derived in \cite{ff}. 
Let $E_i\equiv P^{(3)}_{ii+1}$, that is, $E_i$ acts non-trivially on the 
adjacent spaces $i,\, i+1$. The defining relations of the 
`free-fermions' algebra $\CA$ are:
\bea
& \{ E_i^2, E_{i\pm 1} \} = E_{i\pm 1}\;\;,\;\;\; E_i^3 =E_i &\\
&E_i E_{i\pm 1} E_i =0 \;\;,\;\;\; E_i E_j = E_j E_i \;\;{\rm for}\;\;
|i-j|\geq 2&
\eea
where $\{A,B\}=A B+B A$.
The fourth equation just expresses the fact that 
$E_i$ and $E_j$ commute when they act non-trivially  on disjoint spaces.
The operators  $P^{(1)}$ and $P^{(2)}$ 
form a complete set of projectors on the tensor product space 
$\bC^n\otimes\bC^n$:
\beq
P^{(1)} + P^{(2)}=\bI\;,\;\;  (P^{(1)})^2=P^{(1)}\;,\;\; 
(P^{(2)})^2=P^{(2)}\;,\;\; P^{(1)} P^{(2)}=P^{(2)} P^{(1)}=0
\eeq
where $\bI$ is the identity operator.
The operator $P^{(1)}$ is equal to the square of the operator $P^{(3)}$,
and $P^{(2)}P^{(3)}=P^{(3)}P^{(2)} = 0$.
At $\ga=\pi/2$ the matrix (\ref{rcm}) is equal to the matrix
found in \cite{ff}.

One then checks that $\check{R}$  satisfies the Yang-Baxter equation
\beq
\check{R}_{12}(\la;\ga)  \check{R}_{23}(\la+\mu;\ga)\check{R}_{12}(\mu;\ga) =
\check{R}_{23}(\mu;\ga) \check{R}_{12}(\la+\mu;\ga)
\check{R}_{23}(\la;\ga)\label{ybec}
\eeq
where the parameter $\ga$ is the same for all six factors. The argument
$\ga$ of $\check{R}$ will not be written in  what follows.
The calculation is straightforward if tedious. One can  expand
both sides of the YBE, use elementary trigonometric relations and
the relations satisfied by the $P^{(i)}$'s to conclude. 
Alternatively, setting $q=e^{i\ga}$ and
$y=e^{i\la}$ and multiplying the YBE by $2iqy$,
one can also verify that the YBE is satisfied  for the six values
$q=0,\pm 1,\pm i,\infty$. It is then enough to 
check that the coefficient of $q^1$ vanishes identically to conclude
that the YBE is satisfied for all values of $\la$, $\mu$ and $\ga$. 

It is easy to check that  the  regularity and unitarity properties hold:
\beq
\check{R}(0)=\bI\,\sin(\ga)\;\;,\;\;\; \check{R}(\la) \check{R}(-\la)=\bI\,
\sin(\ga+\la)\sin(\ga-\la)
\eeq

The rational limit of $\check{R}$ is obtained by letting $\la\rightarrow\ga\la$,
dividing by $\sin\ga$ and taking the limit $\ga\rightarrow 0$. These
manipulations conserve all the properties of the $\check{R}$-matrix.
In particular, one obtains
\bea
&\check{R}(\la)=  P^{(1)} +(1-\la) P^{(2)}+ \la P^{(3)}&\\
&\check{R}(0)=\bI\;\;,\;\;\; \check{R}(\la) \check{R}(-\la)=\bI\,(1-\la^2)&
\eea

Explicit representations of the foregoing algebra were given in \cite{ff}.
Let $E^{\af\be}$ be the $n\times n$ matrix with a one at row $\af$ 
and column $\be$ and zeros otherwise. 
Let $n$, $n_1$ and $n_2$ be three integers such that
\beq
2\leq n \;\;,\;\;\; 1\leq n_1\leq n_2\leq n-1 \;\;,\;\;\; n_1+n_2=n
\eeq
and $A$, $B$ be two disjoint sets whose union is the set of basis 
states of $\bC^n$, with card$(A)=n_1$ and card$(B)=n_2$. 
Let also
\bea
P^{(3)}&=&\sum_{a\in A}\sum_{\beta\in B}\left(x_{a\beta} E^{\beta a}\otimes
E^{a\beta} + x_{a\beta}^{-1} E^{a\beta}\otimes E^{\beta a}\right)\\
P^{(1)}&=&(P^{(3)})^2=\sum_{a\in A}\sum_{\beta\in B}\left(E^{\beta\beta}\otimes
E^{a a} + E^{a a}\otimes E^{\beta\beta}\right)\\
P^{(2)}&=&\bI - P^{(1)}=\sum_{a,a^{'}\in A} E^{a a}\otimes
E^{a^{'} a^{'}} + \sum_{\beta,\beta^{'}
\in B} E^{\beta\beta}\otimes E^{\beta^{'} \beta^{'}}
\eea
The $n_1 . n_2$ parameters $x_{a\beta}$ are arbitrary complex numbers. 
Latin indices belong to $A$ while greek indices belong to $B$.
These operators satisfy all the defining relations of the algebra $\CA$.
The restriction 
$n_1\leq n_2$ just avoids a double counting
of distinct models since one has the obvious symmetry $A\leftrightarrow B$. 
For $n_1=1$, $n_2=n-1$, and all the $x_{a\beta}$ equal to each other,
one recovers the  $su(n)$ XX models
found in \cite{mm}. Allowing the twist parameters $x$ to be unequal amounts 
to a multiple deformation of these models. 

It is possible to perform a `gauge' transformation which puts
these models in a Temperley-Lieb form. One obtains 
\bea
&\check{R}^{\rm TL}(\la)=\bI \,\sin(\ga-\la) + (P^{(3)} + P^{\rm TL}_\ga)\,\sin(\la)&\\
&P^{\rm TL}_\ga=\sum_{{a\in A}}\sum_{{\be\in B}}
\left( e^{i\ga}E^{\be\be}\otimes E^{aa}+e^{-i\ga} 
E^{aa}\otimes E^{\be\be} \right)&
\eea
$\check{R}^{{\rm TL}}(\la)$  satisfies  the Yang-Baxter equation.

The transfer matrix is the generating functional of the infinite set 
of conserved quantities.
Its construction in the framework of the Quantum Inverse Scattering 
Method is by now well known \cite{qism2,qism3,fadd}.
The Lax operator on a chain  at site $i$ with inhomogeneity $\mu_i$
is given by:
\beq
L_{0i} (\la) = R_{0i}(\la-\mu_i) = \CP_{0i}  \check{R}_{0i}(\la-\mu_i) 
\eeq
where $\CP$ is the permutation 
operator on $\bC^n\otimes\bC^n$.
The monodromy matrix is a product of Lax operators,
$T(\lambda)= L_{0M}(\la)...L_{01}(\la)\label{mono}$,
where $M$ is the number of sites on the chain and 0 is the auxiliary
space. 
The transfer matrix is the trace of the monodromy matrix over 
the auxiliary space 0: $ \tau (\la)= {\rm Tr}_0 \;\left[\left( L_{0M}...
L_{01}\right)(\la)\right]$.
A set of conserved quantities is then  given by
\beq
H_{p+1} = \left({d^p \ln\tau (\la)\over d\lambda^p}\right)_{\la=0}
\;\;\; , \;\; p\geq 0 \label{cqs}
\eeq 
The  YBE implies the following intertwining relations for the
elements of the monodromy matrix: 
\beq
\check{R}(\la_1-\la_2) \; T(\la_1)\otimes T(\la_2) =  T(\la_2)\otimes T(\la_1)
\;\check{R}(\la_1-\la_2)\label{rtt}
\eeq
Taking the trace over the auxiliary spaces, and using 
the cyclicity property of the trace, one obtains $[\tau(\la_1),\tau(\la_2) ]=0$.
The hamiltonians $H_p$ therefore mutually commute.

The quadratic hamiltonian calculated from (\ref{cqs}) is equal to
\beq
H_2= \frac{1}{\sin\ga}\sum_i \left( P^{(3)}_{ii+1} - P^{(2)}_{ii+1}
\cos\ga \right) \label{h2}
\eeq 
Periodic boundary conditions are implied. 
These  hamiltonians   can be written simply  
in terms of  $su(n)$ hermitian traceless matrices.
For $|x_{a\beta}|=1$ and $\ga$ real the hamiltonians are hermitian.
The rational limit yields: $H_2=\sum_i ( P^{(3)}_{ii+1} - P^{(2)}_{ii+1})$.
It is possible to get a positive  relative sign between $P^{(3)}$ and 
$P^{(2)}$; one lets $\ga\rightarrow\ga+\pi$, and takes the rational limit
for the rational hamiltonian. 

For all the representations of the  operators $P^{(i)}$ given above
one can introduce the `conjugation' matrix  \cite{ff}
\beq
C=\sum_{\beta\in B} E^{\beta\beta} -\sum_{a\in A} E^{aa}\label{cc}
\eeq
and verify that $P^{(2)}=\frac{1}{2}(\bI+C\otimes C)$. 
Dropping the part proportional to the identity in (\ref{h2})
gives the quadratic  XXC hamiltonian. For $n=2$ this is nothing 
but the XXZ model for $su(2)$. 

For all the parameters equal to one parameter,
$x_{a\be}=x\,, \;\forall a\in A\;,\; \forall\be\in B$, 
the linear magnetic-field operators
\beq
H_1^{ab}= \sum_i E_i^{ab}\;\;\; a,b \in A\;\;,\;\;\;
H_1^{\af\be}= \sum_i E_i^{\af\be}\;\;\; \af,\be  \in B
\eeq
commute with the transfer matrix. The proof of \cite{ff} still holds.
The full local symmetry is $su(n_1)\oplus su(n_2)\oplus u(1)$.
One consequence is that one can add   magnetic-field terms
for each symmetry generators, without
spoiling the integrability of the models.

\section{Algebraic Bethe Ansatz}

The diagonalization by algebraic Bethe Ansatz of the XXC models 
is  similar to the one for the   XX models.  We refer the 
reader to \cite{ff,mm} for details and give here the 
results.
I am considering the case $x_{a\be}=x$ to avoid unessential complications which
are peculiar to the higher dimensional models $(n\geq 3)$. 

It is easy to see that the
vector $||1\rangle \equiv |1\rangle\otimes ... \otimes|1\rangle$ 
is an eigenvector of the transfer matrix.
A set of  Bethe Ansatz eigenvectors is then obtained from the
action of certain elements of the monodromy matrix, the $C_{\af}\equiv
T_{1\alpha}$,
and is given by:
\beq
|\lambda_1,..., \lambda_p\rangle =
\sum_{\af_p,..., \af_1} F^{\af_p,..., \af_1}
C_{\af_1}(\lambda_1)...C_{\af_p}(\lambda_p) \; ||1\rangle
\eeq 
where the parameters
$\lambda_i$ and the coefficients $F$ are to be determined.
This Ansatz also vanishes identically if $p>M$; the proof of this fact
is similar to that of \cite{ff}.

One then applies the transfer matrix on the state $|\lambda_1,...,
\lambda_p\rangle$ and with the help of relations (\ref{rtt}) commutes 
it through the $C_\af$'s. 
The corresponding 
eigenvalues of $\tau(\la)$ are
\bea
\Lambda^{((n_1,n_2),M)}(\la;\{ \mu_i\})&=&\left(\prod_{i=1}^M
\sin(\ga-(\la-\mu_i))\right)\prod_{j=1}^p f(\la_j-\la) +  
\prod_{i=1}^M x^{-1}\sin(\la-\mu_i) \nonumber\\
& \times & \Lambda^{(n_2,p)}\prod_{j=1}^p f(\la-\la_j)
+  (n_1-1) \left(\prod_{i=1}^M x\sin(\la-\mu_i)\right)\,\delta_{pM}\label{lamb}
\eea
where $f(\la)=\frac{x \sin(\ga-\la)}{\sin\la}$. 
Here $\Lambda^{(n_2,p)}$ is an eigenvalue of the 
unit-shift operator $\tau^{(n_2,p)}$, 
for a chain of $p$ sites and $n_2$ possible states at each site.
The coefficients $F^{\af_p,..., \af_1}$ are such $F$ is an eigenvector
of $\tau^{(n_2,p)}$ for the eigenvalue $\Lambda^{(n_2,p)}$.
Finally the Bethe Ansatz equations are just
\beq
(-1)^{p-1} \prod_{i=1}^M \frac{x \sin(\ga-(\la_j-\mu_i))} {\sin(\la_j-\mu_i)}=\Lambda^{(n_2,p)}\prod_{k\not=j}^M
\frac{\sin(\ga-(\la_j-\la_k))}{\sin(\ga-(\la_k-\la_j))}\;\; , \; j=1,...,p
\label{baes}
\eeq
Note that these equations imply the vanishing of the residues 
of $\Lambda^{((n_1,n_2),M)}(\la;\{ \mu_i\})$ at the $\la_j$'s. 

The operator $\tau^{(n_2,p)}$ can be written as a product 
of disjoint  permutation cycles. One also  
has $\left(\tau^{(n_2,p)}\right)^p= \bI$ .  
The eigenvalues $\Lambda^{(n_2,p)}$ 
are then
$q^{\rm th}$ roots of unity, where $q$ is a divisor of $p$, 
and are highly degenerate. The degeneracy of an eigenvalue
depends on  $p$ and $n_2$. 

One can perform the above diagonalization procedure over the pseudo-vacuum
$||a\rangle$ ($a\not=1$). The set of eigenvalues is exactly the same
as the one found above and the eigenvectors have the same structure
but form a completely  distinct set, at least for $0\leq p < M$. 
One can also start with any of the vectors $||\be\rangle$, $\be\in B$
and obtain yet other sets of eigenvectors. 
The superscript $n_2$ is replaced by $n_1$ in (\ref{lamb}), 
$\Lambda^{(n_2,p)}$  and  $\tau^{(n_2,p)}$.
These features reflect a large degeneracy of the spectrum. 

The rational limit is obtained by letting $\la_* \rightarrow \ga\la_*$,
$\mu_i \rightarrow \ga\mu_i$,
dividing the eigenvalue (\ref{lamb}) by $(\sin\ga)^M$ and taking the limit 
$\ga\rightarrow 0$. 
For $\ga=\pi/2$ the above results reduce to those of \cite{ff}.
For arbitrary $\ga$ the eigenvalues and the Bethe Ansatz 
equations have the form of those of the $su(2)$ XXZ model, if one
momentarily `forgets' the $\Lambda^{(n_2,p)}$ correction. 
One can also diagonalize the unit-shift operator by algebraic
Bethe Ansatz, since this operator  appears systematically 
as one of the conserved quantities for all the $su(n)$ XXZ spin-chains.
This reflects the underlying symmetry seen above.

A detailed study of the spectrum is desirable. In particular,
the thermodynamic limit of the spectrum would shed some light
on the differences between the XXZ spectrum and the XXC one. 
Taking the logarithm of the Bethe Ansatz equations shows that 
the distribution of integers characterizing the solutions will 
get a contribution from the  $\Lambda^{(n_2,p)}$ correction.
This may result in a denser spectrum. 
Whether this influences the central charge and the conformal 
weights remains to be clarified. 

A feature of the models with $n\geq 3$, 
is the term $\delta_{pM}$ in the expression
of the eigenvalues. This seems to imply a jump in the spectrum of 
$\tau(\la)$. However note that, because of the definition of 
the conserved quantities, the hamiltonians $H_1,..., H_M$ are not 
affected by this $\delta$-term.  For vanishing inhomogeneities, 
the $M^{\rm th}$ power of the eigenvalues of $\tau(0)/(\sin\ga)^M$ 
is equal to one because one has the unit-shift operator for $M$ sites
and $n$ states; one can verify that the $M^{\rm th}$ power of the eigenvalues
(\ref{lamb}) is equal to one, after using equations (\ref{baes}).
The eigenvalues of $H_2$ (at $\mu_i=0$) for the 
trigonometric and rational cases are equal to 
\bea
\frac{d\ln\Lambda}{d\la}|_{\la=0}&=&-M\frac{\cos\ga}{\sin\ga} +\sin\ga\sum_{j=1}^p
\frac{1}{\sin(\la_j)\sin(\ga-\la_j)}\\
\frac{d\ln\Lambda_r}{d\la}|_{\la=0}&=&-M +\sum_{j=1}^p
\frac{1}{\la_j(1-\la_j)}
\eea

Finally, for $\ga\not= \pi/2 + ik\pi$, $k\in\bZ$,
the decorated YBE is not satisfied and it is not possible to couple, 
\`a la Shastry \cite{sh123,ff},
two XXC copies for generic values of $\ga$.


\end{document}